\documentclass[3p,twocolumn]{elsarticle}

\usepackage{amsmath,amssymb,graphicx,bm,nicefrac}
\usepackage{times,txfonts}
\usepackage[normalem]{ulem}

\newcommand{\beqn}{\begin{equation}}
\newcommand{\eeqn}{\end{equation}}
\newcommand{\bea}{\begin{eqnarray}}
\newcommand{\eea}{\end{eqnarray}}

\newcommand{\nmax}{\ensuremath{N_{\rm max}}}

\newcommand{\NNLO}{N$^2$LO}
\newcommand{\NNNLO}{N$^3$LO}

\begin{document}

\title{{\it Ab initio} study of neutron drops with chiral Hamiltonians}

\author[isu]{\mbox{H.D.\ Potter}\corref{cor2}}
\ead{hdpotter@iastate.edu}

\author[tud]{\mbox{S.\ Fischer}\corref{cor1}}
\ead{s.fischer@stud.tu-darmstadt.de}

\author[isu]{\mbox{P.\ Maris}}
\ead{pmaris@iastate.edu}

\author[isu]{\mbox{J.P.\ Vary}}
\ead{jvary@iastate.edu}

\author[tud]{\mbox{S.\ Binder}}
\ead{sven.binder@physik.tu-darmstadt.de}

\author[tud]{\mbox{A.\ Calci}}
\ead{angelo.calci@physik.tu-darmstadt.de}

\author[tud]{\mbox{J.\ Langhammer}}
\ead{joachim.langhammer@physik.tu-darmstadt.de}

\author[tud]{\mbox{R.\ Roth}}
\ead{robert.roth@physik.tu-darmstadt.de}

\cortext[cor1]{Corresponding author}
\cortext[cor2]{Principal corresponding author}

\address[isu]{Department of Physics and Astronomy, Iowa State University, Ames, IA\ 50011, USA}
\address[tud]{Institut f\"ur Kernphysik, Technische Universit\"at Darmstadt, 64289 Darmstadt, Germany}


\begin{abstract}
We report \emph{ab initio} calculations for neutron drops in a 10 MeV external harmonic-oscillator trap using
chiral nucleon-nucleon plus three-nucleon interactions.  We present total binding energies, 
internal energies, radii and odd-even energy differences for neutron numbers $N$ = $2 -18$ using the no-core shell model with and without importance truncation.
Furthermore, we present total binding energies for $N$ = $8, 16, 20, 28, 40, 50$ obtained in a coupled-cluster approach.
Comparisons with quantum Monte Carlo results, where available, using Argonne $v^\prime_8$ with three-nucleon interactions reveal important dependences on the chosen Hamiltonian.
\end{abstract}

\date{\today}


\maketitle

\section{Introduction
\label{sec:introduction}}

There has been significant interest in {\it ab initio} solutions for systems of neutron drops trapped 
in external potentials aimed at providing insights into properties of unstable neutron-rich nuclei 
and neutron star matter \cite{Pudliner:1996,Bogner:2011kp,Gandolfi:2010za,Maris:2013rgq}.  
At the same time, comparisons of neutron drop 
results using different microscopic interactions provide information on the 
isovector part of the nucleon-nucleon (NN) interaction and the $T = 3/2$ component
of the three-nucleon (3N) interaction.  With these goals in mind, we present the first 
{\it ab initio} results for pure neutron systems using chiral NN+3N Hamiltonians 
in an external trap and compare with results previously obtained using high-precision phenomenological NN+3N Hamiltonians.

We adopt no-core configuration interaction methods
(e.g., see Refs.~\cite{Navratil:2000ww,Navratil:2000gs,Barrett:2013nh,Maris:2008ax,Navratil:2009ut,Maris:2009bx,Roth:2009cw,Roth:2011ar,Roth:2011vt,Maris:2011as,Navratil:2011zs})
and coupled-cluster theory~\cite{WlDe05,HaPa08,HaHj12,HaPa14,HaPa10,TaBa08a}
which have advanced rapidly in 
recent years, making it feasible to accurately solve fundamental problems in nuclear
structure and reaction physics.
We follow Refs.~\cite{Bogner:2011kp,Maris:2013rgq} for the configuration interaction approach to trapped neutron drops
in the current application. 

At the same time, significant theoretical advances regarding the underlying Hamiltonians, constructed within chiral effective field theory (EFT), provide a firm foundation for nuclear many-body calculations rooted in QCD \cite{EpHa09,MaEn11}, leading us to adopt a chiral EFT Hamiltonian here.
We also make use of the similarity renormalization group (SRG) approach \cite{Glazek:1993rc,Wegner:1994,Bogner:2007rx,Hergert:2007wp,Bogner:2009bt,Furnstahl:2012fn} that provides a straightforward and flexible framework for consistently evolving (softening) the Hamiltonian and other operators, including 3N interactions~\cite{Roth:2011ar,Jurgenson:2009qs,Roth:2013fqa}.

The goal of this paper is twofold. First, we aim to provide results for neutron drop 
systems in a 10 MeV 
harmonic-oscillator (HO) trap using realistic chiral NN+3N interactions 
with uncertainty estimates where feasible.  Second, we present comparisons between
our results and those of other high-quality NN+3N interactions.  In particular, we compare with
results obtained using the Green's function Monte Carlo (GFMC) and auxiliary field diffusion Monte
Carlo (AFDMC) quantum Monte Carlo (QMC) methods \cite{Gandolfi:2010za,Maris:2013rgq}, where the Argonne $v^\prime_8$ (AV8$^\prime$) NN interaction~\cite{Pudliner:1997ck}
was used in conjunction with the Urbana IX (UIX) 3N interaction~\cite{Pudliner:1997ck} 
and with the Illinois-7 (IL7) 3N interaction~\cite{Pieper:2008}. 

We limit our investigations to a single form of the chiral NN+3N
interaction. That is, we use the chiral NN interaction at \NNNLO\ with 500\,MeV/c cutoff
from Ref.~\cite{Entem:2003ft} together with the chiral 3N interaction at
\NNLO~\cite{Epelbaum:2002vt} in the local form of
Ref.~\cite{Navratil:2007zn} for 500\,MeV/c cutoff with low-energy constants determined entirely in the three-nucleon sector \cite{GaQu09}. 
This is also the Hamiltonian used in Refs.~\cite{Roth:2011ar,Jurgenson:2009qs,Roth:2013fqa,Jurgenson:2010wy,Jurgenson:2013yya,Maris2014zz} for \emph{ab initio} studies of nuclear properties. We evolve this Hamiltonian using the free-space SRG to two representative flow parameters or momentum scales to examine the
scale-dependence of our results. As in the earlier applications, we retain
the induced many-body interactions up to the three-nucleon level and
neglect induced four-nucleon (and beyond) interactions. Depending on whether the initial chiral 3N interaction is included or not we use the term NN+3N-full or NN+3N-induced, respectively, to characterize the SRG-evolved Hamiltonian.

In selected cases, we also compare our results with those obtained using
JISP16 \cite{Shirokov:2003kk,Shirokov:2005bk}, a nonlocal NN potential without 3N interactions.
The neutron drop results with JISP16 have appeared previously in Ref.~\cite{Maris:2013rgq}.

In Section~\ref{sec:back}, we briefly review the formalism and summarize related results from previous work.
The results for our neutron drop observables are presented in Section~\ref{sec:results}. 
Section~\ref{sec:conclusions} summarizes our conclusions and provides perspectives on future efforts.

\section{Theoretical Framework
\label{sec:back}}

We employ {\it ab initio} configuration interaction and coupled-cluster 
methods to solve for the properties of neutron drops.
In the first approach, the no-core shell model (NCSM), we follow 
Refs.~\cite{Navratil:2000ww,Navratil:2000gs,Barrett:2013nh} where, for a given
NN and 3N interaction we diagonalize
the resulting many-body Hamiltonian in a sequence of truncated HO
basis spaces. These basis spaces are characterized by two parameters: $N_{\max}$ specifies
the maximum number of total HO quanta above the lowest allowed HO Slater determinant and 
$\hbar\omega$ specifies the HO energy of the basis. This latter variable is distinct from
the HO energy of the trap which is fixed to be 10 MeV in the present application. 
The goal is to achieve convergence
as indicated by independence from these two basis parameters. 

We also employ an extension of the NCSM, the importance truncated NCSM (IT-NCSM), 
for which we follow Refs.~\cite{Roth:2009cw,Roth:2011ar,Roth:2011vt,Roth:2013fqa} 
where subspaces of the $N_{\max}$-truncated spaces 
are dynamically selected according to an importance measure derived from perturbation theory.   
The IT-NCSM uses this importance measure $\kappa_\nu$ 
for the individual many-body basis states and retains 
only states with $|\kappa_\nu|$ above a threshold $\kappa_{\min}$ in the model space. 
Through a variation
of this threshold and an {\it a posteriori} extrapolation $\kappa_{\min}\rightarrow 0$
the contribution of discarded states is recovered. We use
the sequential update scheme discussed in Refs.~\cite{Roth:2009cw,Roth:2013fqa}, 
which connects to the full NCSM model space and, thus, the exact
NCSM results in the limit of vanishing threshold. 
We recently compared the IT-NCSM with the NCSM in
basis spaces where calculations in both approaches are feasible~\cite{Maris2014zz}.

For neutron drops that exhibit subshell closure we apply the coupled-cluster (CC) method 
which is capable of providing results for heavier systems. 
We use single-reference CC with singles and doubles excitations~\cite{PuBa82}, in which the ground state $|\Psi\rangle$ of a many-body Hamiltonian is parametrized by the exponential ansatz 
\mbox{$| \Psi \rangle = e^{T_1+T_2} \, | \Phi \rangle$},
where $T_n$ are $n$-particle-$n$-hole excitation operators 
acting on a single Slater-determinant reference state $| \Phi \rangle$, which is the Hartree-Fock determinant in our calculations.
Effects of the $T_3$ clusters are included through an \emph{a posteriori} correction to the energy via the $\rm \Lambda CCSD(T)$~\cite{TaBa08a,Taba08b} method.
The underlying single-particle basis is an HO basis truncated in the principal oscillator quantum number \mbox{$e=2n+l \le e_{\rm max}$}. In this work we quote results from $e_{\rm max}$ = 12 model spaces that are sufficiently well converged.
Including explicit 3N interactions into CC calculations results in a significant increase of the computational expense~\cite{HaPa07,RoBi12,BiPi13}. 
To facilitate the calculations, we use the normal-ordered two-body approximation (NO2B)~\cite{HaPa07,RoBi12} to the 3N interaction, which was shown  to be very accurate in calculations of atomic nuclei~\cite{HaPa07,RoBi12,BiPi13}.
In the NO2B approximation, contributions of the 3N interaction are demoted to lower particle ranks through normal-ordering techniques, and the residual normal-ordered three-body operator is discarded~\cite{HaPa07,RoBi12}.
Due to their enormous number, not all of the 3N matrix elements that would be required by the large model spaces employed by the CC method can be included in the normal-ordering procedure. 
For that reason we impose an energy truncation $e_1+e_2+e_3 \le E_{3\max}= 14$ on the 3N matrix elements. We have checked that our results are converged with respect to this truncation.

We adopt the chiral NN+3N interaction described above 
as this Hamiltonian has been applied in a range 
of {\it ab initio} calculations of light and medium-mass nuclei~\cite{Roth:2011ar,Jurgenson:2009qs,Roth:2013fqa,Jurgenson:2010wy,
Jurgenson:2013yya,Maris2014zz,HuLa13}. 
For a detailed discussion of the SRG evolution in the 3N sector adopted here, 
see Ref.~\cite{Roth:2013fqa}.
Unless otherwise specified, we employ SRG-evolved interactions with $\alpha = 0.08~\text{fm}^4$
corresponding to a momentum scale $\lambda_{\text{SRG}}=\alpha^{-1/4}= 1.88\, \text{fm}^{-1}$. 

In our (IT-)NCSM calculations,
the size of the largest feasible model space is constrained by
the total number of required 3N interaction matrix elements as well as
by the number of many-body matrix elements that are computed and stored
for the iterative Lanczos diagonalization algorithm. Through an efficient $JT$-coupled storage scheme and an on-the-fly decoupling during the calculation of the many-body Hamilton matrix \cite{Roth:2011ar,Roth:2013fqa,OrPo13}, the limit arising from handling the 3N matrix elements has been pushed to significantly larger model spaces.

For the full NCSM calculations we employ the MFDn
code~\cite{DBLP:conf/sc/SternbergNYMVSL08,DBLP:journals/procedia/MarisSVNY10,DBLP:conf/europar/AktulgaYNMV12,CPE:CPE3129,Maris:2013}
that is highly optimized for parallel computing.  
In order to exploit 
parallel architectures that include GPUs, 
such as the Titan facility at Oak Ridge National Laboratory, 
we have developed and implemented new
algorithms that significantly speed up the required 
decoupling transformations~\cite{OrPo13}.
The IT-NCSM calculations are performed with a dedicated code that has been developed to accommodate the specific demands of an importance-truncated calculation in a framework optimized for parallel performance. Due to the reduction of the model-space dimension resulting from the importance truncation, typically by two orders of magnitude, the many-neutron Hamiltonian matrix is significantly smaller and the memory needs are drastically reduced. 
The CC calculations are performed based on a highly-efficient angular-momentum coupled implementation \cite{HaPa10}. The most time-consuming part, the calculation of the $\Lambda$CCSD(T) energy correction, is embarrassingly parallel and therefore exhibits a nearly perfect scaling.
The evaluation of the $\Lambda$CCSD(T) energy correction of an $N=50$ neutron drop can be performed within a few thousand CPU hours.

\section{Results
\label{sec:results}}

\begin{figure}[thb]
\includegraphics*[width=3.2in]{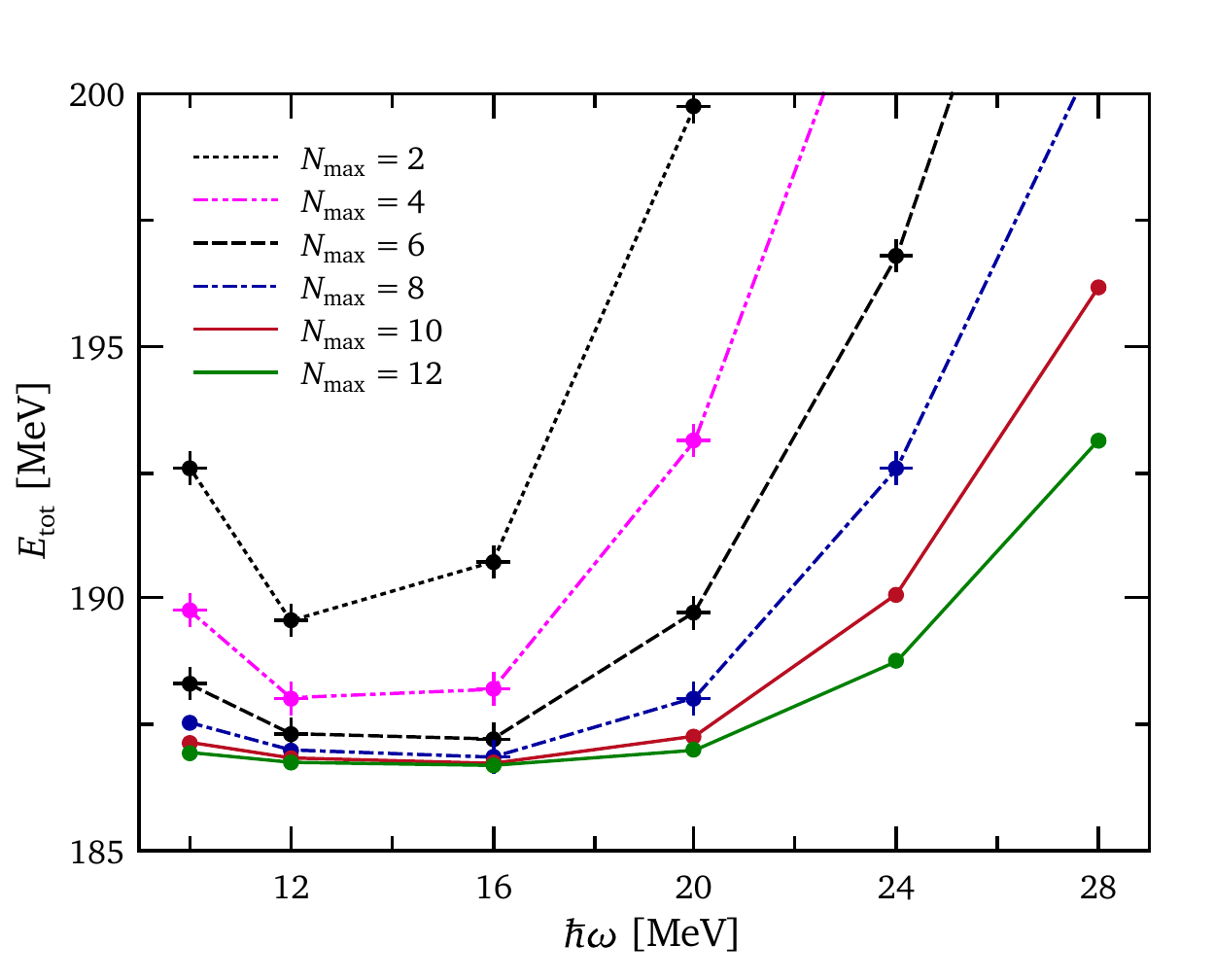}
\caption{(Color online.) Converging sequence of total energies for 10 neutrons in an external
HO potential of 10 MeV with the full chiral NN+3N interaction with 3N matrix elements truncated at $E_{3\max}= 14$. IT-NCSM results are represented by circles, and NCSM results are represented by crosses; results differ by less than the size of the symbols.
\label{fig:convergence_energy}}
\end{figure}

\begin{figure}[thb]
\includegraphics*[width=3.2in]{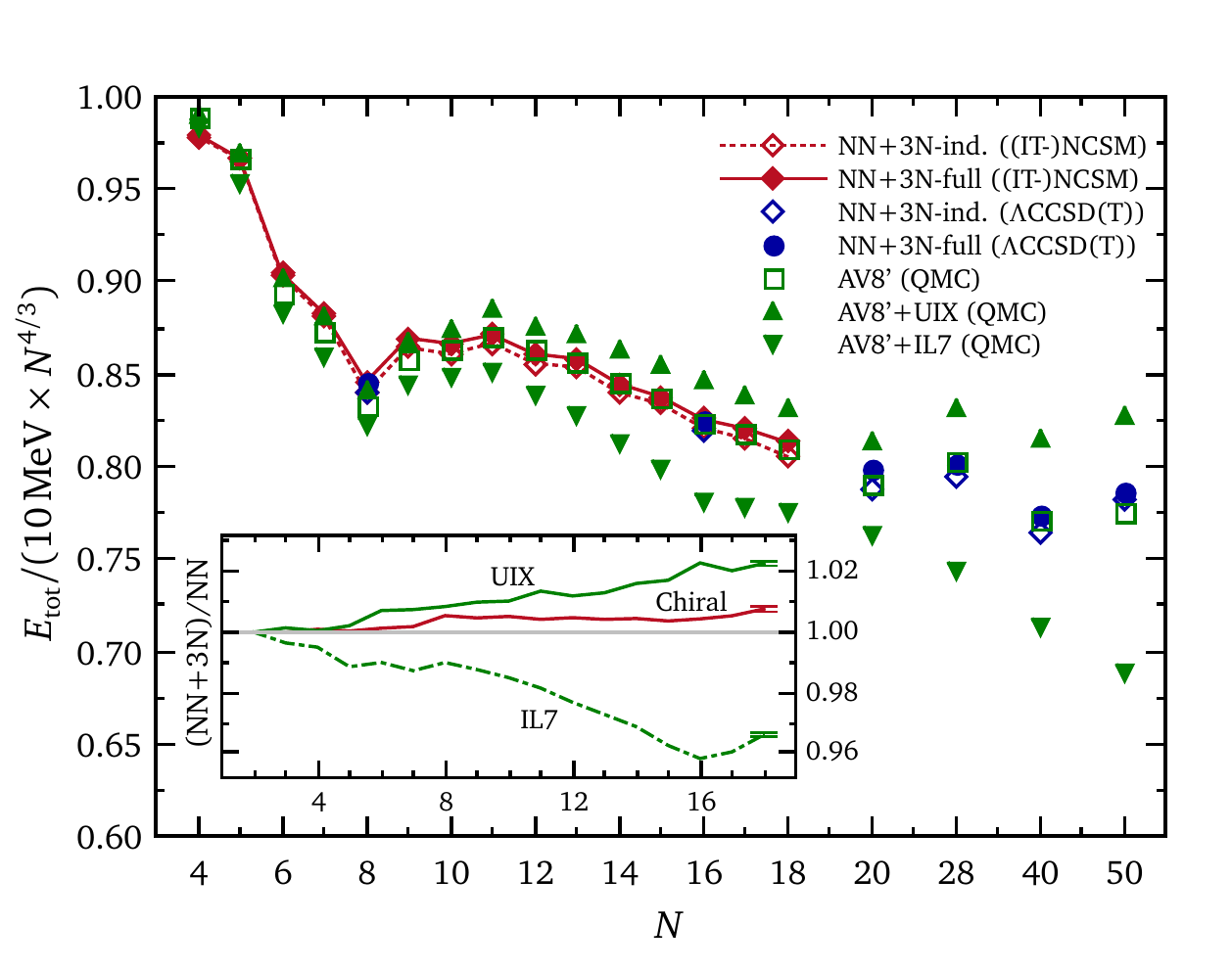}
\caption{(Color online.) Total energy (scaled) of $N$-neutron systems in a 10 MeV HO trap
for different Hamiltonians as a function of $N$. The results for the chiral interactions are obtained in the IT-NCSM or NCSM with the largest accessible $N_{\max}$ (cf. Table \ref{Numerical}) and for closed subshells in $\Lambda$CCSD(T) (cf. Table \ref{NumericalCC}). The results labeled AV8$^\prime$, AV8$^\prime$+UIX and AV8$^\prime$+IL7 are adopted from Ref.~\cite{Maris:2013rgq}.
The inset shows the ratio of the total energies obtained with the initial 3N interaction and without for the largest $N_{\max}$.
Characteristic uncertainties are shown in the inset for the largest $N$ only.
We discuss the uncertainties for our results in the text.
\label{fig:Total_energy_scaled}}
\end{figure}

\renewcommand{\tabcolsep}{3pt}
\begin{table}\footnotesize
\begin{tabular}{c c | r r r | r r }
 & &  \multicolumn{3}{c|}{NCSM} & \multicolumn{2}{c}{IT-NCSM} \\
 \hline
 & & \multicolumn{2}{c}{3N-full} & \multicolumn{1}{c|}{3N-ind.} & \multicolumn{1}{c}{3N-full} & \multicolumn{1}{c}{3N-ind.}\\
$N$ & $J^\pi$ & \multicolumn{1}{c}{$\alpha=0.04$} & \multicolumn{1}{c}{$\alpha=0.08$} & \multicolumn{1}{c|}{$\alpha=0.08$} & \multicolumn{1}{c}{$\alpha = 0.08$} & \multicolumn{1}{c}{$\alpha = 0.08$} \\
\hline
2 & $0^+$ & 23.88(1) & 23.897(1) & 23.897(1) & 23.897(1) & 23.897(1)\\
3 & $3/2^-$ & 45.51(2) & 45.534(4) & 45.532(4) & 45.533(1) & 45.531(1) \\
4 & $0^+$ & 62.17(4) & 62.207(8) & 62.133(7) & 62.205(3) & 62.130(2) \\
5 & $3/2^-$ & 82.65(7) & 82.67(3) & 82.63(3) & 82.651(4) & 82.612(5) \\
6 & $0^+$ & 98.6(1) & 98.62(4) & 98.46(4) & 98.603(5) & 98.443(5) \\
7 & $1/2^-$ & 118.2(1) & 118.24(6) & 117.97(5) & 118.22(2) & 117.95(2) \\
8 & $0^+$ & 135.4(2) & 135.34(7) & 134.42(7) & 135.32(2) & 134.40(2) \\
9 & $5/2^+$ & 162.8(2) & 162.8(1) & 161.84(9) & 162.77(3)  & 161.82(3) \\
10 & $0^+$ & 187.1(6) & 186.8(4) & 185.6(3) & 186.69(4) & 185.49(4) \\
11 & $3/2^+$ & 213.6(7) & 213.3(4) & 212.2(4) & 213.15(5) & 212.03(5) \\
12 & $0^+$ & 237.1(8) & 236.7(5) & 235.3(4) & 236.52(5) & 235.12(5)\\
13 & $5/2^+$ & 263.0(9) & 262.6(6) & 261.2(5) & 262.37(6) & 261.03(6) \\
14 & $0^+$ & 286(1) & 285.3(7) & 283.7(6) & 285.06(7) & 283.47(7) \\
15 & $1/2^+$ & 311(1) & 310.1(7) & 308.6(6) & 309.9(2) & 308.5(2) \\
16 & $0^+$ & 334(1) & 333.1(8) & 331.2(7) & 332.8(3) & 331.0(2) \\
17 & $3/2^+$ & 361(2) & 360(2) & 357(2) & 358.7(3) & 356.3(3) \\
18 & $0^+$    & 386(3) & 385(2) & 381(2) & 383.6(4) & 380.0(3) \\
\end{tabular}
\caption{Comparison of total ground-state energies in units of MeV for $N=2$ 
through $18$ neutrons in a 10 MeV HO trap. We use the SRG-evolved NN+3N-full Hamiltonian, which includes the initial chiral 3N interaction at SRG evolution scales
$\alpha$~=~0.04 and 0.08 fm$^4$, corresponding to momentum scales $\lambda_{\text{SRG}}$ = 2.24 and 1.88 fm$^{-1}$, respectively.
We also present results for the NN+3N induced Hamiltonian without initial chiral 3N at SRG evolution scale $\alpha$~=~0.08~fm$^4$.
Uncertainties, as explained in the text, are quoted in parenthesis for the last
significant figure. 
We present NCSM results at $N_{\rm max} = (14,12,10,8,6)$ for $N = (2,3-4,5-9,10-16,17-18)$. 
For the IT-NCSM we present results at $N_{\rm max} = (14,12,10)$ 
for $N = (2-6,7-14,15-18)$. All results are obtained at basis frequency $\hbar\omega=16$ MeV with 3N matrix elements up through $E_{3\max}= 14$.}
\label{Numerical}
\end{table}

\renewcommand{\tabcolsep}{2pt}
\begin{table}\footnotesize
\begin{tabular}{c | rr rr | rr rr }
 & \multicolumn{8}{c}{$\Lambda$CCSD(T)} \\
\hline
   &  \multicolumn{4}{c|}{3N-full} & \multicolumn{4}{c}{3N-ind.} \\
&  \multicolumn{2}{c}{$\alpha=0.04$} & \multicolumn{2}{c|}{$\alpha=0.08$} &
\multicolumn{2}{c}{$\alpha=0.04$} & \multicolumn{2}{c}{$\alpha=0.08$} \\
$N$ & $E_{\rm tot}$ & $\Delta E$ & $E_{\rm tot}$ & $\Delta E$ & $E_{\rm tot}$ & $\Delta E$ & $E_{\rm tot}$ & $\Delta E$ \\
\hline
8 & 135.1(1)   & -3.7  & 135.3(1)   & -3.0  & 134.2(1)   & -3.6  & 134.4(1)   & -2.9  \\
16 & 332.0(6)   & -11.1  & 332.4(5)   & -8.5  & 329.9(7)   & -10.9  & 330.5(5)   & -8.6  \\
20 & 432.8(5)   & -11.1  & 433.3(3)   & -7.7  & 427.1(4)   & -9.9  & 427.9(2)   & -6.8  \\
28 & 681(1)   & -19.6  & 681(1)   & -14.0  & 674(1)   & -17.8  & 676(1)   & -13.0  \\
40 & 1058(2)   & -25.9  & 1058(1)   & -16.4  & 1043(2)   & -20.9  & 1045(1)   & -12.9  \\
50 & 1449(3)   & -36.6  & 1448(2)   & -23.3  & 1438(5)   & -29.4  & 1442(4)   & -18.3  \\
\end{tabular}
\caption{Total ground-state energies $E_{\text{tot}}$ and correlation energies $\Delta E$ in units of MeV at closed subshells obtained with coupled-cluster theory at the $\Lambda$CCSD(T) level for a 10 MeV HO trap. We use the SRG-evolved NN+3N-full and NN+3N-induced Hamiltonian at SRG evolution scales
$\alpha$ = 0.04 and 0.08 fm$^4$ as in Table \ref{Numerical}. We use Hartree-Fock reference states obtained in an HO single-particle basis with $e_{\max}=12$ and $\hbar\omega=16$ MeV.
Uncertainties are quoted in parenthesis for the last
significant figure and discussed in the text.
}
\label{NumericalCC}
\end{table}

We begin with a demonstration of the ground-state energy convergence with increasing 
basis space truncation $\nmax$ and with a range of HO basis $\hbar\omega$ values as shown 
in Fig.~\ref{fig:convergence_energy} for $N=10$ neutrons using the NCSM and the IT-NCSM. We observe an excellent agreement of the NCSM and the IT-NCSM energies wherever both are available. 
As guaranteed by the variational principle, the results converge
uniformly from above for all values of $\hbar\omega$. The convergence is fastest 
for basis $\hbar\omega$ values slightly above the HO trap strength of 10 MeV as may be expected. 
Since we find the convergence pattern for all other neutron numbers very similar to that shown in 
Fig.~\ref{fig:convergence_energy}, we will quote the lowest energy for fixed neutron number at
the largest $N_{\rm max}$ obtained as our final result for the total ground state energy. From the convergence pattern
in Fig. \ref{fig:convergence_energy} we deduce that it is reasonable to take our uncertainty as the
difference between the quoted energy and the energy at the next smaller 
value of $N_{\rm max}$ at the same $\hbar\omega$ value. 

We then portray our results for the total energy in Fig.~\ref{fig:Total_energy_scaled}
scaled by the Thomas-Fermi $N$-dependence ($N^{4/3}$) and by the HO well strength
following the practice of Refs. \cite{Bogner:2011kp,Gandolfi:2010za,Maris:2013rgq}.
Remarkably, as seen in Fig.~\ref{fig:Total_energy_scaled} as well as in Tables~\ref{Numerical} and \ref{NumericalCC}, our results with chiral Hamiltonians are rather insensitive to the presence or absence
of the initial 3N interaction.  This indicates that the contribution of the chiral 3N interaction 
is very small in the $T = 3/2$ channel.
For comparison, we show our results with the AV8$^\prime$
plus 3N interactions (UIX and IL7) AFDMC results from Refs. \cite{Gandolfi:2010za,Maris:2013rgq} 
in Fig.~\ref{fig:Total_energy_scaled}.
(Ref.~\cite{Maris:2013rgq} provides multiple states with different $J$ for odd neutron drops; in all of our figures with AV8$^\prime$, AV8$^\prime$+UIX or AV8$^\prime$+IL7 we use the states with lowest total energy for comparison with our results.)
To highlight the
difference in the sensitivity to the 3N interaction, we provide an inset in 
Fig.~\ref{fig:Total_energy_scaled} depicting the ratio of the results with the 
initial NN+3N interaction to those with the initial NN interaction alone. For the chiral interactions, 
this ratio fluctuates at a level of less than 1\% over this range of $N$.  For the 
other interactions, it is either an increasing (AV8$^\prime$+UIX) 
or a decreasing (AV8$^\prime$+IL7) function of $N$ through $N=16$, deviating by up to 2\% above or 4\% below, respectively.

We find a pronounced dip in the total energy for the chiral interactions due 
to the expected shell closure at $N=8$ that is similar to dips seen in the results
with the other interactions.   For $N$ $\geq 10$ we observe an apparent coincidence
where the chiral results follow closely those of AV8$^\prime$ without 3N interactions.

We tabulate ground-state energies (unscaled) 
in Table \ref{Numerical} with and without the initial 3N interaction at an SRG evolution parameter of $\alpha=0.08$~fm$^4$ and compare the NCSM and IT-NCSM results.
With the initial 3N interaction we also list our NCSM results at $\alpha=0.04$~fm$^4$, which shows that there is only a weak dependence on the SRG evolution parameter; without the initial 3N interaction we find a similarly weak dependence on the SRG parameter.
Note that the uncertainties at $\alpha=0.08$~fm$^4$ are generally smaller than at $\alpha=0.04$~fm$^4$, as expected.  Furthermore, our results
are consistent between the NCSM and IT-NCSM to within
quoted uncertainties. Note that the IT-NCSM energies are sometimes obtained for larger $N_{\max}$ and are below the NCSM results in those cases, as expected from the variational principle. The uncertainties quoted in the table are estimated from the difference of the total energies in the two largest model spaces. In the case of the IT-NCSM the additional uncertainties from the threshold extrapolation are typically an order of magnitude smaller than the quoted uncertainties.

In Table \ref{NumericalCC} we summarize the total ground-state energies as well as the correlation energies beyond Hartree-Fock obtained from $\Lambda$CCSD(T) calculations for neutrons with closed subshells up to $N=50$. The quoted uncertainties are again estimated from the energy difference in the two largest model spaces. In addition, they include the effects of the $E_{3\max}$-truncation of the 3N matrix elements and of contributions beyond the triply excited clusters in a very conservative estimate. We observe an excellent quantitative agreement of $\Lambda$CCSD(T), IT-NCSM, and NCSM results well within the estimated uncertainties.

As mentioned above, all of these results show that the inclusion of the chiral 3N interaction affects the total energies by less than 1\%, even in the heaviest neutron drops considered here. Furthermore, the dependence on the SRG flow parameter, indicative for the contribution of SRG-induced multi-nucleon interactions beyond the 3N level, is extremely small. Note that the NN+3N-full Hamiltonian adopted here leads to a 
significant flow-parameter dependence in medium-mass nuclei \cite{RoBi12}. This indicates that the main origin of the flow-parameter dependence in symmetric nuclear systems is the $T=1/2$ components of the chiral 3N interaction.

\begin{figure}[t!]
\includegraphics*[width=3.2in]{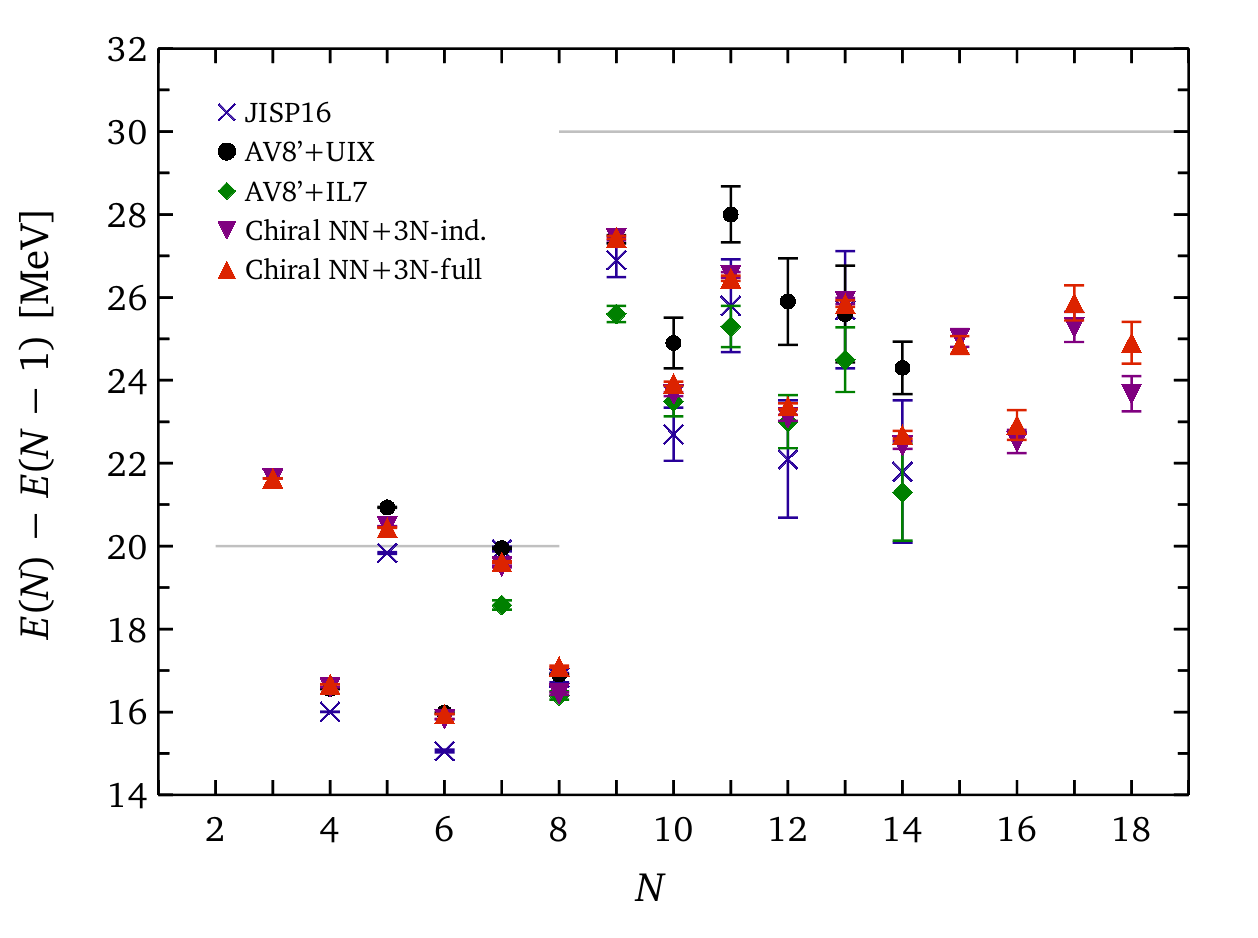}
\includegraphics*[width=3.2in]{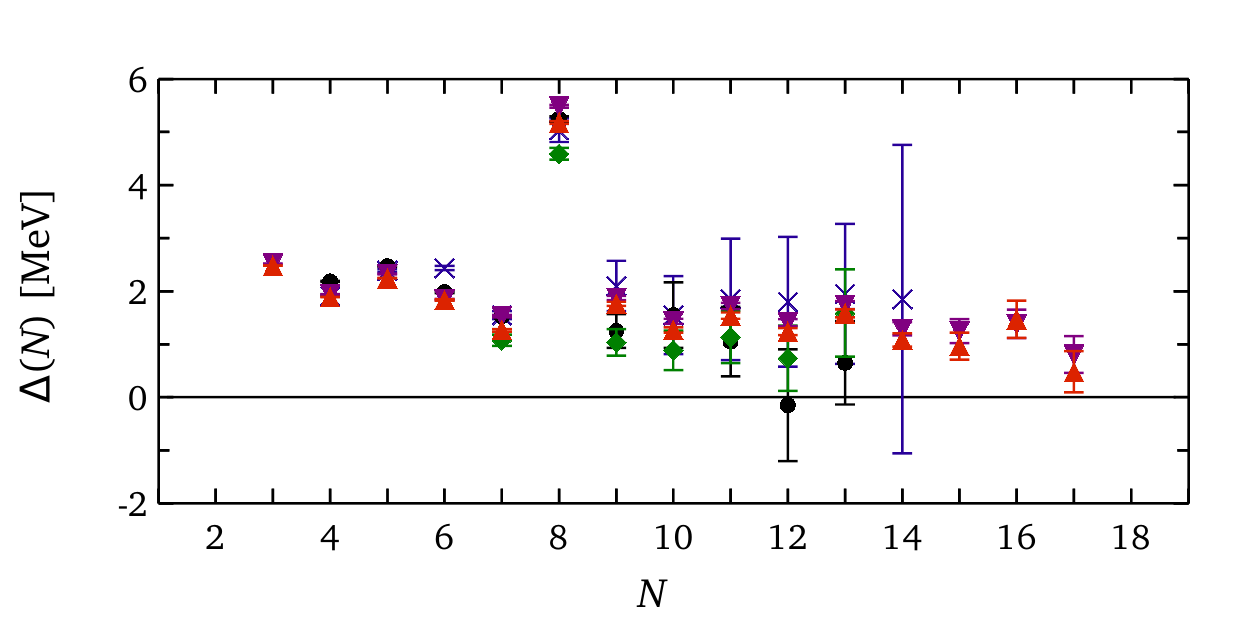}
\caption{(Color online.) Total ground-state energy differences (top) and double differences (bottom) for neutron drops with
neighboring numbers of neutrons for various interactions as indicated in the legend.  
The results for AV8$^\prime$+3N and JISP16 are taken from Ref.~\cite{Maris:2013rgq}; we portray GFMC results for AV8$^\prime$+3N.
\label{diffs}}
\end{figure}

The energy differences between neighboring systems in $N$ are shown in Fig. \ref{diffs}
for several Hamiltonians, including the nonlocal NN interaction JISP16~\cite{Shirokov:2003kk,Shirokov:2005bk}.
Without interactions, we expect the single differences to be simple multiples of the 
HO trap energy as indicated by the solid horizontal lines in the top panel of Fig. \ref{diffs}.
The interactions produce the strong odd-even effect conventionally characterized 
as a ``pairing energy'' effect.
The pairing is more evident in the double differences $\Delta(N)$, defined as
\begin{equation}\Delta(N)=(-1)^{(N-1)}(E(N)-\frac{1}{2}(E(N-1)+E(N+1))),\end{equation}
which are shown in the bottom panel of Fig.~\ref{diffs}.

Overall, we do see the effects of pairing throughout the $p$-shell and the $sd$-shell with the chiral interactions, both with and without the initial 3N interaction.  In the $p$-shell, up to $N=8$, the pairing effects are in rough agreement with the JISP16 results and with the results for AV8$^\prime$+UIX and AV8$^\prime$+IL7.  Above $N=10$, AV8$^\prime$+IL7 and in particular AV8$^\prime$+UIX seem to suggest a smaller pairing energy than we find with the chiral interactions; however, the numerical uncertainties are also significantly larger for these results.

\begin{figure}[thb]
\includegraphics*[width=3.2in]{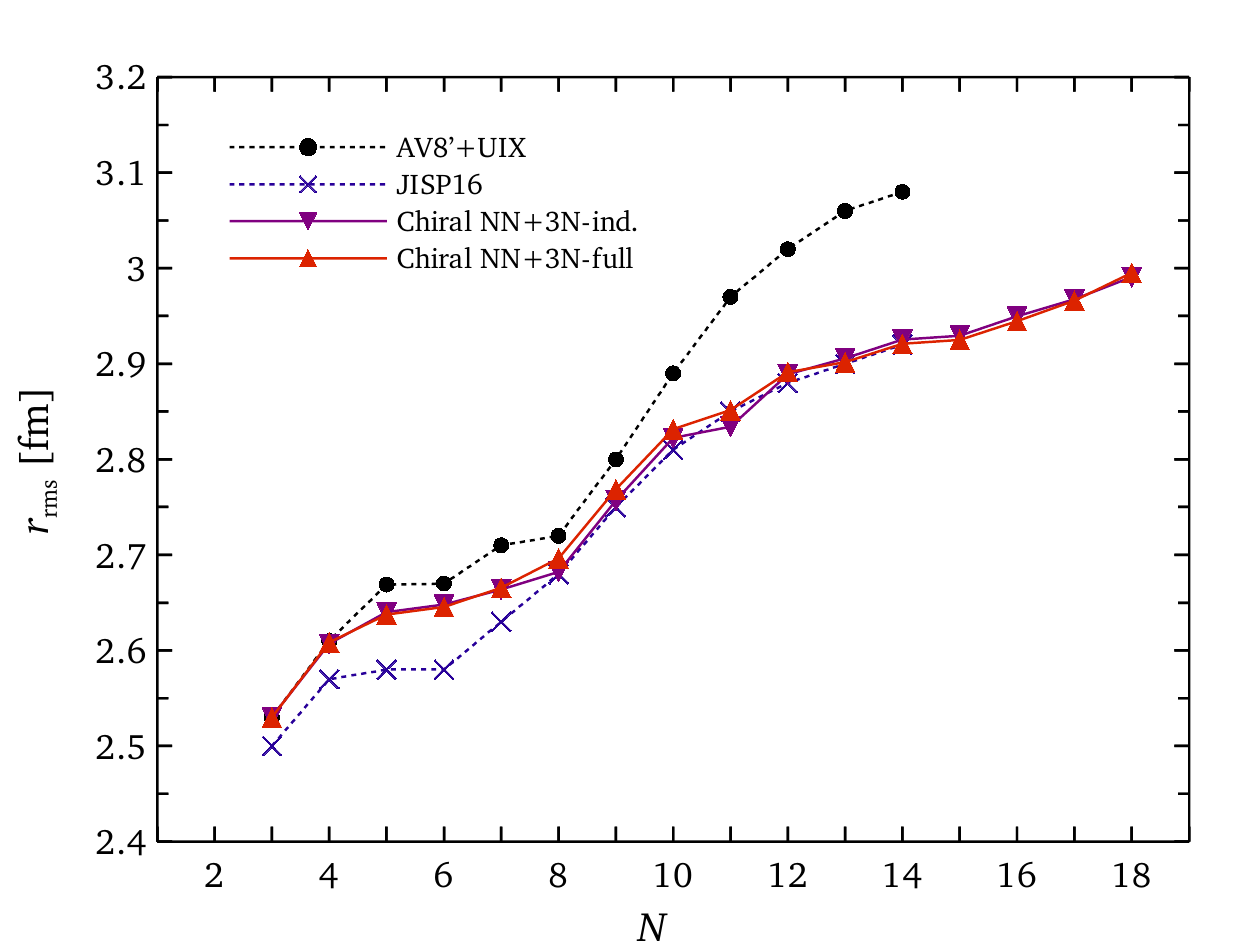}
\caption{
(Color online.) Single-particle rms radii for the lowest energy states of $N=3-18$ systems in a 10 MeV HO trap for various interactions as indicated in the legend. The results for AV8$^\prime$+UIX and JISP16 are taken from Ref.~\cite{Maris:2013rgq} (note Fig.~9 of Ref.~\cite{Maris:2013rgq} shows the rms radii of the states with lowest $J$, which are not always the lowest energy states).
The chiral results with and without full 3N forces are nearly indistinguishable at this scale.
\label{rms_radii}}
\end{figure}

\begin{figure}[thb]
\includegraphics*[width=3.2in]{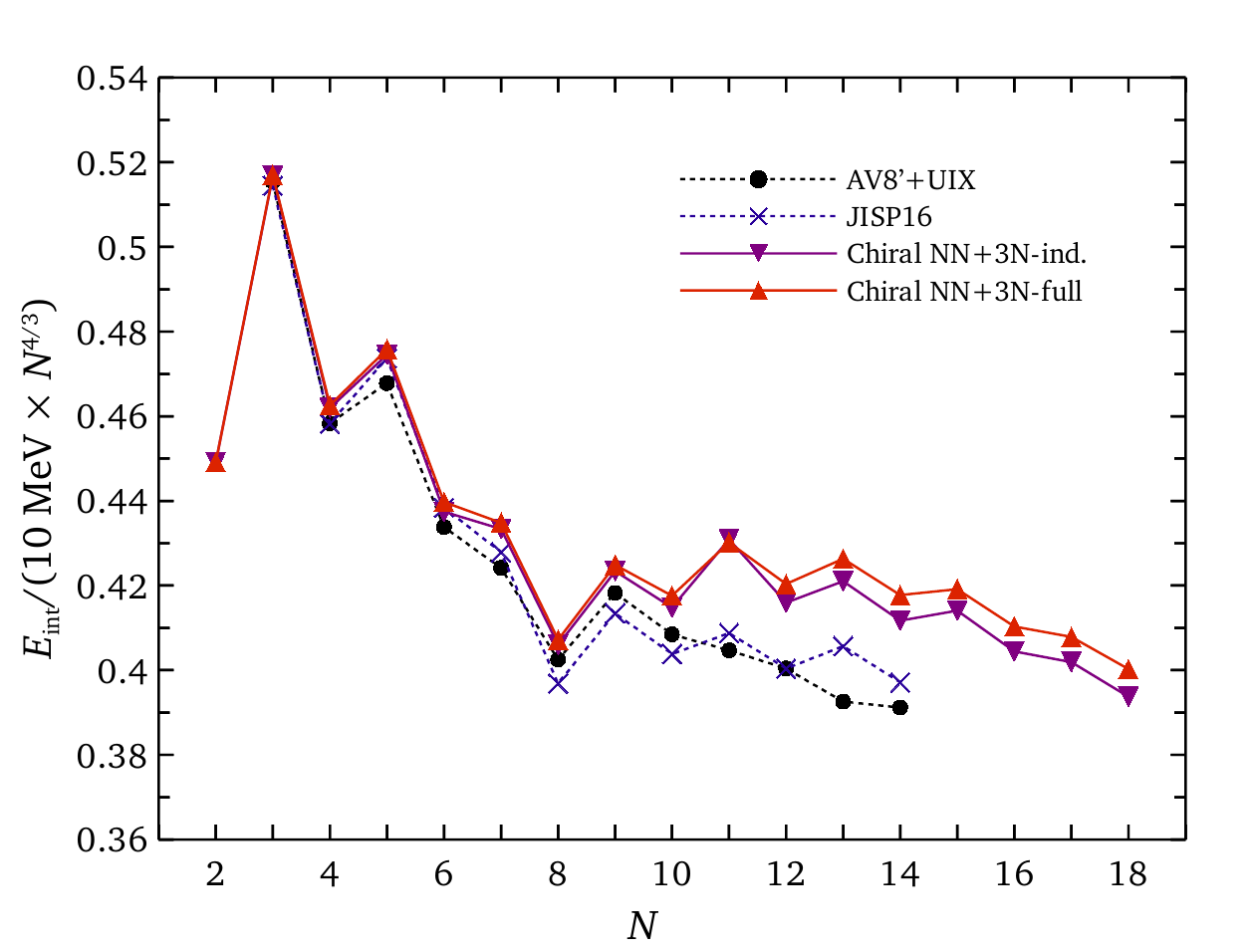}
\caption{(Color online.) Internal energy (scaled) for the lowest-energy states of $N$-neutron systems in a 10 MeV HO trap
as a function of $N$ for various interactions as indicated in the legend. The results for AV8$^\prime$+UIX 
and JISP16 are taken from Ref.~\cite{Maris:2013rgq}.
\label{Internal_E}}
\end{figure}

We present the single-particle root-mean-square (rms) radii in Fig.~\ref{rms_radii} for a selection of Hamiltonians.  Our results for the chiral Hamiltonian show a monotonic increase with $N$, qualitatively similar to the curves with the other interactions.  Above $N=7$ our rms radii follow closely those obtained with JISP16, whereas the radii obtained with AV8$^\prime$+UIX are significantly larger.  It is noteworthy
that the chiral results for the rms radii are again rather insensitive to the presence or absence
of the chiral 3N interaction.

We portray the internal energies (scaled) in Fig.~\ref{Internal_E} which are defined as the total energies of Fig.~\ref{fig:convergence_energy} less the expectation value of the HO trap potential computed with the ground-state wavefunction.
Here one observes the strong odd-even staggering due to pairing, both with the chiral interactions and with JISP16.
AV8$^\prime$+UIX exhibits pairing through $N=9$, but very little for the heavier neutron drops.  We also see, again, the lack of a significant sensitivity of the chiral results to the inclusion of the 3N interaction.

\section{Summary and Conclusions}
\label{sec:conclusions}

We have presented NCSM, IT-NCSM, and CC results for neutron drops in a 10 MeV 
external HO trap using chiral NN+3N interactions. We examined 
total binding energies, odd-even energy differences, internal energies, and radii.  
By comparing with QMC results using AV8$^\prime$ plus 3N interactions we 
found significant dependences on the selected Hamiltonian which should have an impact 
on phenomenological energy-density functionals that may be derived from them.  
Furthermore, we found surprisingly weak contributions in these neutron-drop observables 
from the inclusion of the chiral 3N interaction. Based on systematic trends shown 
in previous neutron-drop investigations 
\cite{Bogner:2011kp,Gandolfi:2010za,Maris:2013rgq}, we anticipate these conclusions will
persist over a range of HO well strengths from 5 MeV to 20 MeV and will follow the trends
seen here for larger values of the neutron number $N$ as well.

Furthermore, as more than one many-body method is applied to the same problem, we 
establish that our NCSM, IT-NCSM and CC results are consistent with each other within the quoted 
uncertainties. We find little sensitivity of these results to the SRG evolution of the chiral interactions
over the range of the SRG flow parameter we investigated.

Following recent practice, we have adopted chiral NN and chiral 3N interactions
that are available up to \NNNLO\ for NN but only up to \NNLO\ for 3N. It is not possible
to estimate the direction or magnitude of changes to expect in our results when
chiral interactions complete through \NNNLO\ become available.  In particular, it will
be very interesting to find out, through detailed calculations,
if the $T=3/2$ components of the 3N interaction complete through \NNNLO\
continue to produce contributions as small as we find them in the present work.

Future work will include the study of excited states and of the sensitivity of observables to the underlying chiral Hamiltonian. Furthermore, the comparison of our {\it ab initio} results to predictions from state-of-the-art energy-density functionals will help constrain the latter in regimes of extreme isospin.

\section*{Acknowledgements}
\label{sec:acknowledgements}
This work was supported in part by the US National Science Foundation
under Grant No. PHY--0904782, the U.S. Department of
Energy (DOE) under Grant Nos.~DE-FG02-87ER40371 and DESC0008485
(SciDAC-3/NUCLEI), by the Deutsche Forschungsgemeinschaft through contract SFB 634, by the Helmholtz International Center for FAIR (HIC for FAIR) within the LOEWE program of the State of Hesse, and the BMBF through contract 06DA7047I. 
This work was supported partially through GAUSTEQ 
(Germany and U.S. Nuclear Theory Exchange Program for QCD Studies of Hadrons and Nuclei)
under contract number \MakeUppercase{DE}-SC0006758. 
A portion of the computational resources were
provided by 
the National Energy Research Scientific Computing Center
(NERSC), a DOE Office of Science User Facility supported by the US DOE Office of Science
under Contract No. DE-AC02-05CH11231, and by 
an
INCITE award, ``Nuclear Structure and Nuclear Reactions'', from the US DOE
Office of Advanced Scientific Computing.  This research also used
resources of the Oak Ridge Leadership Computing Facility at the Oak Ridge National Laboratory,
which is supported by the US DOE Office of Science under Contract No.
DE-AC05-00OR22725. Further resources were provided by the computing 
center of the TU Darmstadt (Lichtenberg), the J\"ulich Supercomputing Centre (JUROPA), and the LOEWE-CSC Frankfurt.
We also wish to thank the ECT* Institute for its hospitality during the 2013 Workshop From Few-Nucleon Forces
to Many-Nucleon Structure and during the 2014 Workshop on Three Body 
Forces: From Matter to Nuclei where portions of this work were completed.


\end{document}